-----------------------------------------------------------------------------

\documentstyle[12pt]{article}

\begin{document}
\topmargin 0pt
\oddsidemargin 7mm
\headheight 0pt
\topskip 0mm

\addtolength{\baselineskip}{0.40\baselineskip}

\hfill SOGANG-HEP 193/94

\hfill October 1994

\begin{center}

\vspace{36pt}
{\large \bf Successive Superalgebraic Truncations
from the Four-Dimensional Maximal Supergravity}

\end{center}

\vspace{36pt}

\begin{center}

Chang-Ho Kim

{\it Department of Physics, Seonam University, Namwon, Chonbuk 590-170, Korea}

\vspace{12pt}

Young-Jai Park$^*$, Kee Yong Kim, and Yongduk Kim

{\it Department of Physics and Basic Science Research Institute, \\
Sogang University, C.P.O. Box 1142, Seoul 100-611, Korea}

\end{center}

\begin{center}
{\bf ABSTRACT}
\end{center}

We study the four-dimensional {\it N}=8 maximal supergravity in the context
of Lie superalgebra SU(8/1).
All possible successive superalgebraic truncations
from four-dimensional {\it N}=8 theory to {\it N}=7, 6, $\cdots$, 1
supergravity theories are systematically realized as sub-superalgebra chains
of SU(8/1) by using the Kac-Dynkin weight techniques.

\vspace{2.5cm}


\vspace{12pt}

\noindent

\vspace{12pt}

\vfill
%\vspace{2cm}
\hrule
\vspace{0.5cm}
\hspace{-0.6cm}$^*$ E-mail address : yjpark@ccs.sogang.ac.kr

\newpage
\begin{center}
{\large \bf I. Introduction}
\end{center}

There have been considerable interests in superalgebras[1,2]
which are relevant to superunifications[3],
nuclear physics[4], supergravities[5], and
superstring theories[6].  Supersymmemtric extensions of
Poincar\'{e} algebra in
{\it D}-dimensional space-time were reviewed,
and their representations (reps)
for the supermultiplets of all known supergravity theories were extensively
searched by Strathdee[7].  This work has been an extremely useful guideline
for
studying supergravity.  Reps of supermultiplets have been obtained within
the
framework of the little algebras of the super-Poincar\'{e} algebra.
Fermionic
generators, which produce supertranslations, are adjoined to the algebra
of the
Poincar\'{e} group.  Cremmer[8] developed the method for consistent
trunctions
by choosing a particular rep of real symplectic metric in order to derive
{\it N}=6,4,2 supergravities from {\it N}=8 in five dimensions.
But this method
is too complicated.

On the other hand, during several years, we have shown that superalgebras
allow a more systematic
analysis for finding the supermultiplets [9,10] of several supergravity
and type-IIB closed superstring theories by using the Kac-Dynkin weight
techniques
of the SU({\it m}/{\it n}) Lie superalgebra[11].
Recently, we have shown that
the massless reps of supermultiplets of the {\it D}=10, {\it N}=2 chiral
supergravity[12] and the {\it D}=4, {\it N}=8 supergravity[13] belong to only
one irreduclble representation (irrep) of the SU(8/1) superalgebra using
the Kac-Dynkin method[14].

In this paper, we show that the succecive superalgebraic trunctions from the
{\it D}=4, {\it N}=8 maximal supergravity[13] to {\it N}=7, 6, $\cdots$, 1
theories can be systematically realized as sub-superalgebra chains
of SU(8/1) Lie
superalgebra by using projection matrices[15].  In Sec. II, we
briefly recapitulate the mathematical structure of the SU(8/1) superalgebra
related to the {\it D}=4, {\it N}=8 maximal supergravity.
In Sec. III, we explicitly
show that all possible supermultiplets of {\it D}=4, {\it N}=7, 6, $\cdots$,
1 theories can be systematically obtained from
this maximal supergravity by successive superalgebraic
truncations.  The last section contains conclusions.

\vspace{1cm}
\newpage
\begin{center}
{\large \bf II. SU(8/1) Superalgebra with the Kac-Dynkin Method}
\end{center}

The Kac-Dynkin diagram of the SU(8/1) Lie superalgebra is

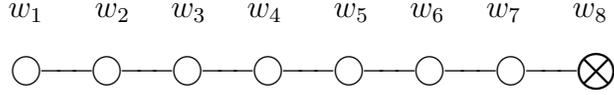
\begin{eqnarray}
w_1~~~~~w_2~~~~ w_3~~~~ w_4~~~~~ w_5~~~~ w_6~~~~ w_7~~~~~ w_8~ \nonumber \\
\bigcirc \!\!-\!\!\!-\!\!\!-\!\!\bigcirc \!\!-\!\!\!-\!\!\!-\!\!
\bigcirc \!\!-\!\!\!-\!\!\!-\!\!\bigcirc \!\!-\!\!\!-\!\!\!-\!\!
\bigcirc \!\!-\!\!\!-\!\!\!-\!\!\bigcirc \!\!-\!\!\!-\!\!\!-\!\!
\bigcirc \!\!-\!\!\!-\!\!\!-\!\! \bigotimes \nonumber
\end{eqnarray}

\noindent
where the set $(w_1~w_2~\cdots~w_8)$ characterizes the heighest weight vector
of an irrep[1,2]. A weight component $w_i~(i \neq 8)$ should be a nonnative
integer, while $w_8$ can be any complex number. The seven white nodes in the
Kac-Dynkin diagram denote the simple even roots
$\alpha_i~(i=1,~2,~\cdots,~7)$,
which constitute SU(8) subalgebra, while the last node denotes the simple odd
root $\beta_8$.

The corresponding graded Cartan matrix is

\vspace{0.5cm}
\begin{equation}
\left [
\begin{array}{rrrrrrrr}
2  & -1 & 0  & 0  & 0  & 0  & 0  & 0  \\
-1 & 2  & -1 & 0  & 0  & 0  & 0  & 0  \\
0  & -1 & 2  & -1 & 0  & 0  & 0  & 0  \\
0  & 0  & -1 & 2  & -1 & 0  & 0  & 0  \\
0  & 0  & 0  & -1 & 2  & -1 & 0  & 0  \\
0  & 0  & 0  & 0  & -1 & 2  & -1 & 0  \\
0  & 0  & 0  & 0  & 0  & -1 & 2  & -1 \\
0  & 0  & 0  & 0  & 0  & 0  & -1 & 0
\end{array}
\right ]
 .
\end{equation}
\vspace{0.5cm}

\noindent
Each simple even root $\alpha_i$ corresponds to the {\it i}-th column of the
graded Cartan matrix, while the simple odd root $\beta_8$ corresponds to
the last column of the graded Cartan matrix.
Then, the positive and negative simple roots $\alpha_i^\pm$ and $\beta_8^\pm$
are directly read from the graded Cartan matrix as follows;

\begin{equation}
\begin{array}{cccccccccc}
\alpha_1^\pm & = & (\pm2 & \mp1 &  0   & ~~0 & ~~0 &  0   &  0   & 0),  \\
\alpha_2^\pm & = & (\mp1 & \pm2 & \mp1 & ~~0 & ~~0 &  0   &  0   & 0),  \\
             & \vdots \\
\alpha_7^\pm & = & (0    &  0   &  0   & ~~0 & ~~0 & \mp1 & \pm2 & \mp1), \\
\beta_8^\pm  & = & (0    &  0   &  0   & ~~0 & ~~0 &  0   & \mp1 & 0).
\end{array}
\end{equation}

\noindent
On the other hand, other odd roots are easily obtained by

\begin{equation}
\beta_i^\pm = [\alpha_i^\pm ,~ \beta_{i+1}^\pm],~~~i=1,~2,~\cdots,~7.
\end{equation}

\noindent
Note that the action by an odd root $\beta_i^\pm$ alternates a bosonic
floor with a fermionic one.

The fundamental rep of SU(8/1) is (1~0~0~0~0~0~0~0), which is
$[({\bf 8,1})_F ~\oplus~({\bf 1,1})_B]$ in the SU(8)$\otimes$U(1) bosonic
subalgebra basis where the subscripts {\it F} and {\it B} stand for fermionic
and bosonic degrees of freedom, respectively, as follows:
\\
\begin{equation}
\begin{array}{lccl}
                 & (1~0~0~0~0~0~0~0) \\
\\
\mid \mbox{gnd}> & (1~0~0~0~0~0~0~0) & = & ({\bf 8},~{\bf 1}) \\
                 & \Downarrow \beta_1^- \\
\mid \mbox{1st}> & (0~0~0~0~0~0~0~1) & = & ({\bf 1},~{\bf 1}). \\
\end{array}
\end{equation}
\\
\noindent
In contrast to the usual Lie algebra SU({\it n}), the complex conjugate
rep of the fundamental rep is given by $(0~0~0~0~0~0~0~-1) =
[({\bf 1,1})_B~ \oplus~({\bf \overline {8},1})_F]$:
\\
\begin{equation}
\begin{array}{lccl}
                 & (0~0~0~0~0~0~0~-1) \\
\\
\mid \mbox{gnd}> & (0~0~0~0~0~0~0~-1) & = & ({\bf 1},~{\bf 1}) \\
                 & \Downarrow \beta_8^- \\
\mid \mbox{1st}> & (0~0~0~0~0~0~1~-1) & = & ({\bf \overline 8},~{\bf 1}). \\
\end{array}
\end{equation}
\\
\noindent
Note that the last component of the highest weight vector of SU(8/1) relating
with the simple odd root can be any {\it complex} number.
Similar to the usual Lie
algebras, we have the following relation from the tensor product of the above
two reps in Eqs.(4) and (5):
\\
\begin{equation}
(1~0~0~0~0~0~0~0)\otimes(0~0~0~0~0~0~0~-1) = (1~0~0~0~0~0~0~-1)
\oplus(0~0~0~0~0~0~0~0).
\end{equation}

\noindent
Then, one can easily recognize that the even and odd roots generate
the adjoint
rep $(1~0~0~0~0~0~0~-1)$ of SU(8/1) as follows:
\\
\begin{equation}
\begin{array}{llcl}
                 & (1~0~0~0~0~0~0~-1) & \\
\\
\mid \mbox{gnd}> & (1~0~0~0~0~0~0~-1) & = & \beta_i^+ \\
\\
\mid \mbox{1st}> & (1~0~0~0~0~0~1~-1) & = & \mbox{SU(8)} \\
                 & (0~0~0~0~0~0~0~0) & = & \mbox{U(1)} \\
\\
\mid \mbox{2nd}> & (0~0~0~0~0~0~1~0) & = & \beta_i^-. \\
\end{array}
\end{equation}
\\

Now let us consider the irreps of SU(8/1) superalgebra.
There are two types of irreps,
which are {\it typical} and {\it atypical} [1,11,16].  All atypical reps of
SU(8/1) are characterized by
the last component of the highest weight which corresponds to the last node
of the Kac-Dynkin diagram.  The atypicality condition[11] is given by

\begin{equation}
w_8 = - \sum_{j=i}^{7} w_j + i - 8,~~~1\leq i\leq 8.
\label{line1}
\end{equation}

\noindent
Since an odd root $\beta_i^\pm$ string is terminated in the full weight
system for the case of atypical rep, which $w_8$ satisfies Eq.(8)
for a specific
{\it i}, the atypical reps generally have not equal bosonic and fermionic
degrees of freedom.

On the other hand, typical reps of SU(8/1) consist of nine floors due to the
existances of the eight odd roots, and have equal bosonic and fermionic
degrees of freedom.  The lowest dimensional typical rep is
$(0~0~0~0~0~0~0~w_8) = [{\bf 128}_B \oplus {\bf 128}_F]$
for $w_8 \neq$ $0,~-1,~-2,~-3,~-4,~-5,~-6,$ and $-7$ such as

\begin{equation}
\begin{array}{llcl}
                 & (0~0~0~0~0~0~0~w_8) \\
\\
\mid \mbox{gnd}> & (0~0~0~0~0~0~0~w_8) & = & {\bf 1} \\
                 & ~~~~~~~~~ \Downarrow \beta_8^- \\
\mid \mbox{1st}> & (0~0~0~0~0~0~1~w_8) & = & {\bf \overline 8} \\
                 & ~~~~~~~~~ \Downarrow \beta_7^- \\
\mid \mbox{2nd}> & (0~0~0~0~0~1~0~w_8+1) & = & {\bf \overline {28}} \\
                 & ~~~~~~~~~ \Downarrow \beta_6^- \\
\mid \mbox{3rd}> & (0~0~0~0~1~0~0~w_8+2) & = & {\bf \overline {56}} \\
                 & ~~~~~~~~~ \Downarrow \beta_5^- \\
\mid \mbox{4th}> & (0~0~0~1~0~0~0~w_8+3) & = & {\bf 70} \\
                 & ~~~~~~~~~ \Downarrow \beta_4^- \\
\mid \mbox{5th}> & (0~0~1~0~0~0~0~w_8+4) & = & {\bf 56} \\
                 & ~~~~~~~~~ \Downarrow \beta_3^- \\
\mid \mbox{6th}> & (0~1~0~0~0~0~0~w_8+5) & = & {\bf 28} \\
                 & ~~~~~~~~~ \Downarrow \beta_2^- \\
\mid \mbox{7th}> & (1~0~0~0~0~0~0~w_8+6) & = & {\bf 8} \\
                 & ~~~~~~~~~ \Downarrow \beta_1^- \\
\mid \mbox{8th}> & (0~0~0~0~0~0~0~w_8+7) & = & {\bf 1}.
\end{array}
\end{equation}
\\
\noindent
Particularly taking $w_8=-\frac{7}{2}$, the weight system in Eq.(9) shows the
{\it typical} and {\it real} property.  By usig these properties, we have
already shown that the typical rep $(0~0~0~0~0~0~0~-\frac{7}{2})$ is
beautifully
identified with the supermultiplets of the {\it D}=4, {\it N}=8 supergravity
and {\it D}=10, {\it N}=2 chiral supergravity[12]. Note that the rep with
$w_8=0$ in Eq.(9) is reduced to a singlet rep (0~0~0~0~0~0~0~0)
because the odd root $\beta_8^-$ string is terminated due to the atypicality
condition (8). One can also generate the atypical rep $(0~0~0~0~0~0~0~-1)$,
which is the complex conjugate rep of the fundamental rep, in Eq.(5) from
Eq.(9) since the $\beta_7^-$ string is broken for $w_8=-1$.

The next higher dimensional typical rep is (1 0 0 0 0 0 0 $w_8$) =
[${\bf 1024}_B \oplus {\bf 1024}_F$] for $w_8 \neq$ 0, $-1,~-2,~-3,~-4,~-5,
-6,$ and $-8$. Note that the fundamental rep with $w_8 = 0$ in Eq.(4)
and the adjoint
rep with $w_8=-1$ in Eq.(7) are atypical because the $\beta_2^-$ and
$\beta_8^-$ strings are terminated from the typical rep $(1~0~0~0~0~0~0~w_8)$,
respectively.

\newpage
\begin{center}
{\large \bf III. Successive Superalgebraic Truncations
from {\it N}=8 to {\it N}=7,6,$\cdots$,1 in {\it D}=4}
\end{center}

\vspace{1cm}

The massless reps of supermultiplets for {\it D}=4, {\it N}=8 supergravity[7,
13] have the algebraic structure of SU(8)$\otimes$SO(2), where SU(8) is the
supersymmetry algebra and {\it D}=4 light-cone symmetry is SO(2)$\approx$U(1),
{\it i.e.}, the helicity quantum number is given by the eigenvalue
of this U(1) generator.  Since the
bosonic subalgebra of SU(8/1) is SU(8)$\otimes$U(1), we have recently shown
that the well-known {\it D}=4, {\it N}=8 supermultiplets correspond to only
one {\it real}
typical irrep with $w_8=-\frac{7}{2}$ in Eq.(9), which is

\begin{equation}
(0~0~0~0~0~0~0~-\frac{7}{2}) = [{\bf 1}_B \oplus {\bf \overline{8}}_F
                             \oplus {\bf \overline{28}}_B \oplus {\bf
                             \overline{56}}_F \oplus {\bf 70}_B \oplus
                             {\bf 56}_F \oplus {\bf 28}_B \oplus {\bf 8}_F
                             \oplus {\bf 1}_B].
\end{equation}

\noindent
Note that the helicity state of each multiplet crucially depends upon
the condition {\it w}$_8 = -\frac{7}{2}$, which guarantees equal
degrees of freedom for bosons and fermions.
The regular branching SU({\it n}/1)$\rightarrow$SU({\it n})$\otimes$U(1) is
attained by removing the last node from the Kac-Dynkin diagram of
SU({\it n}/1),
and the corresponding projection matrix {\bf P$_1$}({\it n}) is given by

\begin{equation}
{\bf \mbox{P}_1}(n) = \left[
  \begin{array}{cccccc}
    1      & 0      & \cdots & 0      & | & 1 \\
    0      & 1      & 0      & \vdots & | & 2 \\
    \vdots & 0      & \ddots & 0      & | & \vdots \\
    \vdots & \vdots & 0      & 1      & | & n-1 \\
    0      & \cdots & \cdots & 0      & | & n
  \end{array}
                      \right]
 .
\end{equation}

\noindent
By letting {\bf P$_1$}(8) act on the whole weight system of SU(8/1),
and normalizing the U(1) eigenvalues by $-14$, one gets the desired
supermultiplets
as follows:

\begin{equation}
\begin{array}{lcll}
\mbox{floor}     & \mbox{SU(8/1)}               & \mbox{field}
& \mbox{helicity} \\
\\
\mid \mbox{gnd}> & (0~0~0~0~0~0~0~-\frac{7}{2}) & e_\mu^a
& +2 \\
\\
\mid \mbox{1st}> & (0~0~0~0~0~0~1~-\frac{7}{2}) & \overline 8 \Psi_\mu
& +\frac{3}{2} \\
\\
\mid \mbox{2nd}> & (0~0~0~0~0~1~0~-\frac{5}{2}) & \overline {28} A_\mu
& +1 \\
\\
\mid \mbox{3rd}> & (0~0~0~0~1~0~0~-\frac{3}{2}) & \overline {56} \lambda
& +\frac{1}{2} \\
\\
\mid \mbox{4th}> & (0~0~0~1~0~0~0~-\frac{1}{2}) & 70\phi
& ~0 \\
\\
\mid \mbox{5th}> & (0~0~1~0~0~0~0~+\frac{1}{2}) & 56 \lambda
& -\frac{1}{2} \\
\\
\mid \mbox{6th}> & (0~1~0~0~0~0~0~+\frac{3}{2}) & 28 A_\mu
& -1 \\
\\
\mid \mbox{7th}> & (1~0~0~0~0~0~0~+\frac{5}{2}) & 8 \Psi_\mu
& -\frac{3}{2} \\
\\
\mid \mbox{8th}> & (0~0~0~0~0~0~0~+\frac{7}{2}) & e_\mu^a
& -2.
\end{array}
\end{equation}
\\

Now let us consider successive superalgebraic truncations from the {\it D}=4,
{\it N}=8 maximal theory to the {\it N}=7,~6,~$\cdots$,~1 supergravities.
One
must carefully remove extra gravitino multiplets in a consistent manner to
guarantee the existence of the {\it N}=7,~6,~$\cdots$,~1 theories.  We can
easily show that superalgebraic truncations is systematically understood
as a regular branching chain of SU(8/1):

\begin{equation}
\overbrace{\mbox{SU}(8/1)}^{N=8} \rightarrow \overbrace{\mbox{SU}(7/1) \otimes
\mbox{U}(1)}^{N=7} \rightarrow \overbrace{\mbox{SU}(6/1) \otimes
[\mbox{U}(1)]^2}^{N=6} \rightarrow \cdots \rightarrow
\overbrace{[\mbox{U}(1)]^8}^{N=1}
\end{equation}

\noindent
This branching chain is systematically attained by successively removing the
first node from the corresponding Kac-Dynkin diagrams.  The U(1) supercharge
assignment for the branching pattern
SU({\it N}/1) $\rightarrow$ SU({\it N}$-1$/1)$\otimes$U(1) for {\it N}
$\geq$ 2
is given by

\begin{equation}
\mbox{U}(1) = diag.(N-2,\overbrace{-1,-1,\cdots,-1}^{N-\mbox{terms}})
\end{equation}

\noindent
since Eq.(14) must satisfy the supertraceless condition[11].  The
corresponding projection matrix {\bf P$_2$}({\it n}) is given by

\vspace{1cm}
\begin{equation}
{\bf \mbox{P}_2}(n) = \left[
  \begin{array}{ccccccc}
    0      & \cdots & \cdots & \cdots & 0      & | & n-2 \\
    1      & 0      & \vdots & \vdots & \vdots & | & n-3 \\
    0      & 1      & 0      & \vdots & \vdots & | & \vdots \\
    \vdots & 0      & \ddots & 0      & \vdots & | & ~1 \\
    \vdots & \vdots & 0      & 1      & 0      & | & ~0 \\
    0      & \cdots & \cdots & 0      & 1      & | & -1 \\
  \end{array}
                   \right]
 .
\end{equation}\\
Then we are ready to analyse successive superalgebraic trunctions.

\vspace{1cm}
\begin{center}
{\bf 3.1 {\it N} = 7 Supergravity}
\end{center}

First, let us consider the consistent superalgebraic truncation from the
{\it D}=4, {\it N}=8 to the {\it D}=4, {\it N}=7 theory.  By allowing the
matrix {\bf P$_2$}(8) to act on the right hand side of the SU(8/1) weight
system in Eq.(12) that describes of the supermultiplets of the {\it D}=4,
{\it N}=8 theory, one easily
gets the two desired reducible SU(7/1)$\otimes$U$^a$(1) reps for {\it N}=7
as follows:

\begin{equation}
(0~0~0~0~0~0~0~-\frac{7}{2}) \rightarrow (0~0~0~0~0~0~-\frac{7}{2})
(\frac{7}{2})
\oplus(0~0~0~0~0~0~-\frac{5}{2})(-\frac{7}{2}).
\end{equation}

\noindent
Since the complex conjugate rep of $(0~0~0~0~0~0~-\frac{7}{2})(\frac{7}{2})$
is
$(0~0~0~0~0~0~-\frac{5}{2})(-\frac{7}{2})$, the two combined reps in Eq.(16)
still maintain the desired real property although the truncation is occured.

Then one can easily recognize that the typical rep
$(0~0~0~0~0~0~-\frac{7}{2})$
of SU(7/1) with the U$^a$(1) quantum number $h^a = \frac{7}{2}$ is
$[{\bf 64}_B \oplus
{\bf 64}_F]_\uparrow$, which contains the half of the
supermultiplets of the {\it N}=7 theory from the branching rule
SU(7/1)$\rightarrow \mbox{SU}(7) \otimes \mbox{U}^b(1)$ by using
the projection matrix {\bf P$_1$}(7) as follows:

\begin{equation}
\begin{array}{lcll}
\mbox{floor}     & \mbox{SU(7/1)}             & \mbox{field}
& \mbox{helicity} \\
\\
\mid \mbox{gnd}> & (0~0~0~0~0~0~-\frac{7}{2}) & e_\mu^a
& +2 \\
\\
\mid \mbox{1st}> & (0~0~0~0~0~1~-\frac{7}{2}) & \overline 7 \Psi_\mu
& +\frac{3}{2} \\
\\
\mid \mbox{2nd}> & (0~0~0~0~1~0~-\frac{5}{2}) & \overline {21} A_\mu
& +1 \\
\\
\mid \mbox{3rd}> & (0~0~0~1~0~0~-\frac{3}{2}) & \overline {35} \lambda
& +\frac{1}{2} \\
\\
\mid \mbox{4th}> & (0~0~1~0~0~0~-\frac{1}{2}) & 35 \phi
& ~0 \\
\\
\mid \mbox{5th}> & (0~1~0~0~0~0~+\frac{1}{2}) & 21 \lambda
& -\frac{1}{2} \\
\\
\mid \mbox{6th}> & (1~0~0~0~0~0~+\frac{3}{2}) & 7 A_\mu
& -1 \\
\\
\mid \mbox{7th}> & (0~0~0~0~0~0~+\frac{5}{2}) & \Psi_\mu
& -\frac{3}{2}, \\
\end{array}
\end{equation}

\noindent
where the desired helicity quantum number {\it h} is given by

\begin{equation}
h = -\frac{1}{84} h^a -\frac{1}{12} h^b.
\end{equation}

\noindent
The remaining half supermultiplets are identified with
the rep $(0~0~0~0~0~0~-\frac{5}{2}) = [{\bf 64}_B \oplus
{\bf 64}_F]_\downarrow$
with $h^a = -\frac{7}{2}$ as follows:

\begin{equation}
\begin{array}{lcll}
\mbox{floor}     & \mbox{SU(7/1)}             & \mbox{field}
& \mbox{helicity} \\
\\
\mid \mbox{gnd}> & (0~0~0~0~0~0~-\frac{5}{2}) & \Psi_\mu
& +\frac{3}{2} \\
\\
\mid \mbox{1st}> & (0~0~0~0~0~1~-\frac{5}{2}) & \overline {7} A_\mu
& +1 \\
\\
\mid \mbox{2nd}> & (0~0~0~0~1~0~-\frac{3}{2}) & \overline {21} \lambda
& +\frac{1}{2} \\
\\
\mid \mbox{3rd}> & (0~0~0~1~0~0~-\frac{1}{2}) & \overline {35} \phi
& ~0 \\
\\
\mid \mbox{4th}> & (0~0~1~0~0~0~+\frac{1}{2}) & 35 \lambda
& -\frac{1}{2} \\
\\
\mid \mbox{5th}> & (0~1~0~0~0~0~+\frac{3}{2}) & 21 A_\mu
& -1 \\
\\
\mid \mbox{6th}> & (1~0~0~0~0~0~+\frac{5}{2}) & 7 \Psi_\mu
& -\frac{3}{2} \\
\\
\mid \mbox{7th}> & (0~0~0~0~0~0~+\frac{7}{2}) & e_\mu^a
& -2. \\
\end{array}
\end{equation}

\noindent
Note that the upper (lower) arrow indicates that the multiplet contains the
graviton mode having the positive (negative) helicity.
Therefore, the well-known supergravity multiplets $[{\bf 128}_B
\oplus {\bf 128}_F]$
of the {\it D}=4, {\it N}=7 theory[7] given by

\begin{equation}
\begin{array}{cl}
 & [{\bf 1}_2 \oplus {\bf \overline {7}}_{3/2} \oplus {\bf \overline {21}}_1
  \oplus {\bf \overline {35}}_{1/2} \oplus {\bf 35}_0 \oplus {\bf 21}_{-1/2}
                     \oplus {\bf 7}_{-1} \oplus {\bf 1}_{-3/2}] \\
\\
 & ~~~~\oplus~[{\bf 1}_{3/2} \oplus {\bf \overline{7}}_1
                                  \oplus {\bf \overline {21}}_{1/2}
       \oplus {\bf \overline {35}}_0 \oplus
                               {\bf 35}_{-1/2} \oplus {\bf 21}_{-1}
       \oplus {\bf 7}_{-3/2} \oplus {\bf 1}_{-2}],
\end{array}
\end{equation}\\

\noindent
where the states are represented by their SU(7) dimensionality and
U(1) helicity, are exactly identified with two combined irreps of SU(7/1)
in Eqs.(17) and (19).  Note that each ground floor in Eqs.(17) and (19)
originates from the ground floor and the first floor of the SU(8/1) rep
in Eq.(12), respectively.

\vspace{1cm}
\begin{center}
{\bf 3.2 {\it N}=6 Supergravity}
\end{center}

By using the projection matrix {\bf P$_2$}(7) to act on the right hand
side of
the SU(7/1) weight system in Eq.(16), one gets the following four
reducible SU(6/1)$ \otimes $U$^c$(1) reps which are appropriate to {\it N}=6
theory:

\begin{equation}
(0~0~0~0~0~0~-\frac{7}{2}) \rightarrow (0~0~0~0~0~-\frac{7}{2})(\frac{7}{2})
\oplus (0~0~0~0~0~-\frac{5}{2})(-\frac{5}{2}), \\
\end{equation}

\begin{equation}
(0~0~0~0~0~0~-\frac{5}{2}) \rightarrow (0~0~0~0~0~-\frac{5}{2})(\frac{5}{2})
\oplus (0~0~0~0~0~-\frac{3}{2})(-\frac{7}{2}). \\
\end{equation}

\noindent
Then, we can easily show that the typical rep $(0~0~0~0~0~-\frac{7}{2})
= [{\bf 32}_B \oplus {\bf 32}_F]_\uparrow$
with $h^a = h^c = \frac{7}{2}$ contains the half of the supermultiplets of
the {\it N}=6 theory
from the branching rule SU(6/1)$\rightarrow \mbox{SU}(6) \otimes
\mbox{U}^d(1)$ by using
{\bf P$_1$}(6) as follows:

\begin{equation}
\begin{array}{lcll}
\mbox{floor}     & \mbox{SU(6/1)}           & \mbox{field}
& \mbox{helicity} \\
\\
\mid \mbox{gnd}> & (0~0~0~0~0~-\frac{7}{2}) & e_\mu^a
& +2 \\
\\
\mid \mbox{1st}> & (0~0~0~0~1~-\frac{7}{2}) & \overline 6 \Psi_\mu
& +\frac{3}{2} \\
\\
\mid \mbox{2nd}> & (0~0~0~1~0~-\frac{5}{2}) & \overline {15} A_\mu
& +1 \\
\\
\mid \mbox{3rd}> & (0~0~1~0~0~-\frac{3}{2}) & 20 \lambda
& +\frac{1}{2} \\
\\
\mid \mbox{4th}> & (0~1~0~0~0~-\frac{1}{2}) & 15 \phi
& ~0 \\
\\
\mid \mbox{5th}> & (1~0~0~0~0~+\frac{1}{2}) & 6 \lambda
& -\frac{1}{2} \\
\\
\mid \mbox{6th}> & (0~0~0~0~0~+\frac{3}{2}) & A_\mu
& -1, \\
\end{array}
\end{equation}

\noindent
where the desired helicity quantum number {\it h} is

\begin{equation}
h~=~-\frac{1}{84} h^a -\frac{1}{60} h^c -\frac{1}{10} h^d.
\end{equation}

And the other half of the supermultiplets are in the complex conjugate rep
$(0~0~0~0~0~-\frac{3}{2}) = [{\bf 32}_B \oplus {\bf 32}_F]_\downarrow$
in Eq.(21) with the quantum numbers $h^a = h^c = -\frac{7}{2}$, which
is the graviton multiplet with the helicity $-2$ as follows:

\begin{equation}
\begin{array}{lcll}
\mbox{floor}     & \mbox{SU(6/1)}           & \mbox{field}
& \mbox{helicity} \\
\\
\mid \mbox{gnd}> & (0~0~0~0~0~-\frac{3}{2}) & A_\mu
& +1 \\
\\
\mid \mbox{1st}> & (0~0~0~0~1~-\frac{3}{2}) & \overline {6} \lambda
& +\frac{1}{2} \\
\\
\mid \mbox{2nd}> & (0~0~0~1~0~-\frac{1}{2}) & \overline {15} \phi
& ~0 \\
\\
\mid \mbox{3rd}> & (0~0~1~0~0~+\frac{1}{2}) & 20 \lambda
& -\frac{1}{2} \\
\\
\mid \mbox{4th}> & (0~1~0~0~0~+\frac{3}{2}) & 15 A_\mu
& -1 \\
\\
\mid \mbox{5th}> & (1~0~0~0~0~+\frac{5}{2}) & 6 \Psi_\mu
& -\frac{3}{2} \\
\\
\mid \mbox{6th}> & (0~0~0~0~0~+\frac{7}{2}) & e_\mu^a
 & -2. \\
\end{array}
\end{equation}

\noindent
Note that the rest two typical reps  $(0~0~0~0~0~-\frac{5}{2})$ carrying
U$^c(1) = \pm \frac{5}{2}$ are extra gravitino multiplets, which should be
removed for consistency in the {\it N} = 6 theory. As a result, the desired
supergravity multiplets $[{\bf 64}_B\oplus {\bf 64}_F]$ of the
{\it D}=4, {\it N}=6 theory, which is given by

\begin{equation}
\begin{array}{l}
[{\bf 1}_2 \oplus {\bf \overline {6}}_{3/2} \oplus {\bf \overline {15}}_1
\oplus {\bf 20}_{1/2} \oplus {\bf 15}_0 \oplus {\bf 6}_{-1/2}
\oplus {\bf 1}_{-1}] \\
\\
{}~~~~\oplus~[{\bf 1}_1 \oplus {\bf \overline {6}}_{1/2}
\oplus {\bf \overline {15}}_0 \oplus {\bf 20}_{-1/2} \oplus {\bf 15}_{-1}
\oplus {\bf 6}_{-3/2} \oplus {\bf 1}_{-2}],
\end{array}
\end{equation}

\noindent
are exactly identified with two irreps of SU(6/1) in Eqs.(23) and (25).

\vspace{1cm}
\begin{center}
{\bf 3.3 {\it N} = 5 Supergravity}
\end{center}

Next, let us carry out superalgebraic truncations from the {\it N}=6
to the {\it N}=5
theory.  By using the projection matrix {\bf P$_2$}(6),
one easily gets SU(5/1)$\otimes$U$^e$(1) reducible reps
for {\it N}=5 as follows:

\begin{equation}
(0~0~0~0~0~-\frac{7}{2}) \rightarrow
(0~0~0~0~-\frac{7}{2})(\frac{7}{2})\oplus(0~0~0~0~-\frac{5}{2})(-\frac{3}{2})
\end{equation}

\begin{equation}
(0~0~0~0~0~-\frac{3}{2}) \rightarrow
(0~0~0~0~-\frac{3}{2})(\frac{3}{2})\oplus(0~0~0~0~-\frac{1}{2})(-\frac{7}{2})
\end{equation}

\noindent
The typical rep $(0~0~0~0~-\frac{7}{2}) = [{\bf 16}_B \oplus {\bf
16}_F]_\uparrow$ with $h^a = h^c = h^e = \frac{7}{2}$  is relavant
to the graviton multiplet
for the {\it N}=5 theory when SU(5/1) $\rightarrow$ SU(4/1) $\otimes$ U$^f$(1)
as follows

\begin{equation}
\begin{array}{lcll}
\mbox{floor}     & \mbox{SU(5/1)}         & \mbox{field}
& \mbox{helicity} \\
\\
\mid \mbox{gnd}> & (0~0~0~0~-\frac{7}{2}) & e_\mu^a
& +2 \\
\\
\mid \mbox{1st}> & (0~0~0~1~-\frac{7}{2}) & \overline 5 \Psi_\mu
& +\frac{3}{2} \\
\\
\mid \mbox{2nd}> & (0~0~1~0~-\frac{5}{2}) & \overline {10} A_\mu & +1 \\
\\
\mid \mbox{3rd}> & (0~1~0~0~-\frac{3}{2}) & 10 \lambda
& +\frac{1}{2} \\
\\
\mid \mbox{4th}> & (1~0~0~0~-\frac{1}{2}) & 5 \phi               & ~0 \\
\\
\mid \mbox{5th}> & (0~0~0~0~+\frac{1}{2}) & \lambda
& -\frac{1}{2}, \\
\end{array}
\end{equation}

\noindent
where the helicity quantum number {\it h} is

\begin{equation}
h~=~-\frac{1}{84} h^a -\frac{1}{60} h^c -\frac{1}{40} h^e -\frac{1}{8} h^f.
\end{equation}

The remaining half of the supermultiplets are identified with
$(0~0~0~0~-\frac{1}{2}) = [{\bf 16}_B \oplus {\bf 16}_F]_\downarrow$ with
$h^a = h^c = h^e = -\frac{7}{2}$ as follows

\begin{equation}
\begin{array}{lcll}
\mbox{floor}     & \mbox{SU(5/1)}         & \mbox{field}
& \mbox{helicity} \\
\\
\mid \mbox{gnd}> & (0~0~0~0~-\frac{1}{2}) & \lambda
& +\frac{1}{2} \\
\\
\mid \mbox{1st}> & (0~0~0~1~-\frac{1}{2}) & \overline {5} \phi     & ~0 \\
\\
\mid \mbox{2nd}> & (0~0~1~0~+\frac{1}{2}) & \overline {10} \lambda
& -\frac{1}{2} \\
\\
\mid \mbox{3rd}> & (0~1~0~0~+\frac{3}{2}) & 10 A_\mu               & -1 \\
\\
\mid \mbox{4th}> & (1~0~0~0~+\frac{5}{2}) & 5 \Psi_\mu
& -\frac{3}{2} \\
\\
\mid \mbox{5th}> & (0~0~0~0~+\frac{7}{2}) & e_\mu^a                & -2. \\
\end{array}
\end{equation}
\\
\noindent
Therefore, the well-known supergravity multiplets $[{\bf 32}_B \oplus
{\bf 32}_F]$ of the {\it N}=5 theory, which is given by \\
\begin{equation}
[{\bf 1}_2 \oplus {\bf \overline {5}}_{3/2} \oplus {\bf \overline {10}}_1
\oplus {\bf 10}_{1/2} \oplus {\bf 5}_0 \oplus {\bf 1}_{-1/2}]
\oplus~[{\bf 1}_{1/2} \oplus {\bf \overline {5}}_0
\oplus {\bf \overline {10}}_{-1/2} \oplus {\bf 10}_{-1}
\oplus {\bf 5}_{-3/2} \oplus {\bf 1}_{-2}],
\end{equation}

\noindent
are exactly identified with two irreps of SU(5/1) in Eqs.(29) and (31).
The other two typical reps  $(0~0~0~0~-\frac{5}{2})$ and
$(0~0~0~0~-\frac{3}{2})$ in Eqs.(27) and (28) are extra gravitino
multiplets having the maximal helicity
$\pm \frac{3}{2}$.

\vspace{1cm}
\begin{center}
{\bf 3.4 {\it N} = 4 Supergravity}
\end{center}

The next stage of truncation is obtained from the branching rule SU(5/1)
$\rightarrow$
SU(4/1) $\otimes$ U$^g$(1) by using {\bf P$_2$}(5) as follows

\begin{equation}
\begin{array}{ccc}

(0~0~0~0~-\frac{7}{2})~ \rightarrow   &
                       & (0~0~0~-\frac{7}{2})(\frac{7}{2}) \oplus
(0~0~0~-\frac{5}{2})(-\frac{1}{2}), \\
\\
(0~0~0~0~-\frac{1}{2})~ \rightarrow   &
                      & (0~0~0~-\frac{1}{2})(\frac{1}{2}) \oplus
(0~0~0~+\frac{1}{2})(-\frac{7}{2}). \\
\end{array}
\end{equation}
\\
\noindent
Then, the typical rep $(0~0~0~-\frac{7}{2}) = [{\bf 16}_B
\oplus {\bf 16}_F]_\uparrow$
with $h^a = h^c = h^e = h^g =\frac{7}{2}$ contains
the half of the supermultiplets of the {\it N}=4 theory as follows:

\begin{equation}
\begin{array}{lcll}
\mbox{floor}     & \mbox{SU(4/1)}       & \mbox{field}
& \mbox{helicity} \\
\\
\mid \mbox{gnd}> & (0~0~0~-\frac{7}{2}) & e_\mu^a              & +2 \\
\\
\mid \mbox{1st}> & (0~0~1~-\frac{7}{2}) & \overline 4 \Psi_\mu
& +\frac{3}{2} \\
\\
\mid \mbox{2nd}> & (0~1~0~-\frac{5}{2}) & 6 A_\mu              & +1 \\
\\
\mid \mbox{3rd}> & (1~0~0~-\frac{3}{2}) & 4 \lambda
& +\frac{1}{2} \\
\\
\mid \mbox{4th}> & (0~0~0~-\frac{1}{2}) & \phi                 & ~0, \\
\end{array}
\end{equation}
\\
\noindent
where the helicity quantum number is given by

\begin{equation}
h~=~-\frac{1}{84} h^a -\frac{1}{60} h^c -\frac{1}{40} h^e -\frac{1}{24} h^g
-\frac{1}{6} h^h.
\end{equation}

The complex conjugate rep $(0~0~0~+\frac{1}{2}) = [{\bf 16}_B
\oplus {\bf 16}_F]_\downarrow$
with $h^a = h^c = h^e = h^g =-\frac{7}{2}$ of Eq.(34)
contains the remaining half of the supermultiplets as follows:

\begin{equation}
\begin{array}{lcll}
\mbox{floor}     & \mbox{SU(4/1)}       & \mbox{field}
& \mbox{helicity} \\
\\
\mid \mbox{gnd}> & (0~0~0~+\frac{1}{2}) & \phi                  & ~0 \\
\\
\mid \mbox{1st}> & (0~0~1~+\frac{1}{2}) & \overline {4} \lambda
& -\frac{1}{2} \\
\\
\mid \mbox{2nd}> & (0~1~0~+\frac{3}{2}) & 6 A_\mu               & -1 \\
\\
\mid \mbox{3rd}> & (1~0~0~+\frac{5}{2}) & 4 \Psi_\mu
& -\frac{3}{2} \\
\\
\mid \mbox{4th}> & (0~0~0~+\frac{7}{2}) & e_\mu^a               & -2. \\
\end{array}
\end{equation}
\\
\noindent
The other two typical reps  $(0~0~0~-\frac{5}{2})$ with $h^a = h^c = h^e =
\frac{7}{2}$, and $h^g = -\frac{1}{2}$, and $(0~0~0~-\frac{1}{2})$
with $h^a = h^c
= h^e = -\frac{7}{2}$, and $h^g = +\frac{1}{2}$ stand for extra gravitino
multiplets.

It seems appropriate to comment on two extra gravitino multiplets in
{\it N}=5
theory. By using the matrix {\bf P$_2$}(5), from the gravitino reps
of {\it N}=5 theory
we can easily generate the supermultiplets $[{\bf 8}_B \oplus {\bf 8}_F]$
of {\it D}=4, {\it N}=4 Yang-Mills theory, which are

\begin{equation}
[{\bf 1}_1 \oplus {\bf \overline {4}}_{1/2} \oplus {\bf 6}_0 \oplus
{\bf 4}_{-1/2} \oplus {\bf 1}_{-1}],
\end{equation}

\noindent
in terms of SU(4/1)$\otimes$U$^g$(1) reps as follows:

\begin{equation}
\begin{array}{c}
(0~0~0~0~-\frac{5}{2}) \rightarrow (0~0~0~-\frac{5}{2})(\frac{5}{2})
\oplus (0~0~0~-\frac{3}{2})(-\frac{3}{2}), \\
\\
(0~0~0~0~-\frac{3}{2}) \rightarrow (0~0~0~-\frac{3}{2})(\frac{3}{2})
\oplus (0~0~0~-\frac{1}{2})(-\frac{5}{2}).
\end{array}
\end{equation}

\noindent
Each {\it real} rep $(0~0~0~-\frac{3}{2})$ in Eq.(38) is exactly an Yang-Mills
supermultiplet which has the contents as follows

\begin{equation}
\begin{array}{lcll}
\mbox{floor}     & \mbox{SU(4/1)}       & \mbox{field}
& \mbox{helicity} \\
\\
\mid \mbox{gnd}> & (0~0~0~-\frac{3}{2}) & A_\mu              & +1 \\
\\
\mid \mbox{1st}> & (0~0~1~-\frac{3}{2}) & \overline 4 \lambda
& +\frac{1}{2} \\
\\
\mid \mbox{2nd}> & (0~1~0~-\frac{1}{2}) & 6 \phi              & ~0 \\
\\
\mid \mbox{3rd}> & (1~0~0~+\frac{1}{2}) & 4 \lambda
& -\frac{1}{2} \\
\\
\mid \mbox{4th}> & (0~0~0~+\frac{3}{2}) & A_\mu               & -1. \\
\end{array}
\end{equation}

\noindent
As a result, each extra gravitino multiplet of {\it N}=5 theory splits
into an
gravitino and an Yang-Mills multiplets of {\it N}=4 theory in
SU(4/1)$\otimes $U$^g$(1) sub-superalgebra basis.

\vspace{1cm}
\begin{center}
{\bf 3.5 {\it N}=3 Supergravity} \\
\end{center}

Next, let us carry out superalgebraic truncation from the {\it N}=4
to the {\it N}=3
theory.  By using {\bf P$_2$}(4), we get {\it N}=3 multiplets from the
branching rule SU(4/1)$\rightarrow$ SU(3/1)$\otimes$U$^i$(1) as follows:

\begin{equation}
\begin{array}{ccl}

(0~0~0~-\frac{7}{2}) & \rightarrow & (0~0~-\frac{7}{2})(+\frac{7}{2})
                             \oplus(0~0~-\frac{5}{2})(+\frac{1}{2}), \\
\\
(0~0~0~+\frac{1}{2}) & \rightarrow & (0~0~+\frac{1}{2})(-\frac{1}{2})
                              \oplus(0~0~+\frac{3}{2})(-\frac{7}{2}). \\
\end{array}
\end{equation}

\noindent
The typical rep $(0~0~-\frac{7}{2})$ with $h^a=h^c=h^e=h^g=h^i= \frac{7}{2}$
is  $[{\bf 4}_B \oplus {\bf 4}_F]_\uparrow$, which contains the half of the
supermultiplets of the {\it N}=3 theory as follows:
\\
\begin{equation}
\begin{array}{lcll}
\mbox{floor}     & \mbox{SU(3/1)}     & \mbox{field}
& \mbox{helicity} \\
\\
\mid \mbox{gnd}> & (0~0~-\frac{7}{2}) & e_\mu^a              & +2 \\
\\
\mid \mbox{1st}> & (0~1~-\frac{7}{2}) & \overline 3 \Psi_\mu
& +\frac{3}{2} \\
\\
\mid \mbox{2nd}> & (1~0~-\frac{5}{2}) & 3 A_\mu              & +1 \\
\\
\mid \mbox{3rd}> & (0~0~-\frac{3}{2}) & \lambda
& +\frac{1}{2}, \\
\end{array}
\end{equation}

\noindent
where the helicity quantum number is given by

\begin{equation}
h = -\frac{1}{84} h^a -\frac{1}{60} h^c -\frac{1}{40} h^e
-\frac{1}{24} h^g -\frac{1}{12} h^i -\frac{1}{4} h^j.
\end{equation}

The complex conjugate rep $(0~0~+\frac{3}{2})$ with $h^a=h^c=h^e=h^g=h^i=
-\frac{7}{2}$ is $[{\bf 4}_B \oplus {\bf 4}_F]_\downarrow$, which contains
the
remaining half of the supermultiplets as follows:

\begin{equation}
\begin{array}{lcll}
\mbox{floor}     & \mbox{SU(3/1)}     & \mbox{field}  & \mbox{helicity} \\
\\
\mid \mbox{gnd}> & (0~0~+\frac{3}{2}) & \lambda       & -\frac{1}{2} \\
\\
\mid \mbox{1st}> & (0~1~+\frac{3}{2}) & \overline 4 A_\mu & -1 \\
\\
\mid \mbox{2nd}> & (1~0~+\frac{5}{2}) & 4 \Psi_\mu    & -\frac{3}{2} \\
\\
\mid \mbox{3rd}> & (0~0~+\frac{7}{2}) & e_\mu^a           & -2. \\
\end{array}
\end{equation}

As a result, the desired supergravity multiplets $[{\bf 8}_B
\oplus {\bf 8}_F]$
of the {\it D}=4, {\it N}=3 theory, which is given by

\begin{equation}
[{\bf 1}_2 \oplus {\bf \overline {3}}_{3/2} \oplus {\bf 3}_1
\oplus {\bf 1}_{1/2} ]
\oplus [{\bf 1}_{-1/2} \oplus {\bf \overline {3}}_{-1}
\oplus {\bf 3}_{-3/2} \oplus {\bf 1}_{-2}],
\end{equation}

\noindent
are identified with the two irreps in Eqs.(42) and (44).

On the other hand, the Yang-Mills multiplet  of {\it N}=4 goes into Yang-Mills
multiplets $[{\bf 8}_B \oplus {\bf 8}_F]$ of {\it N} =3 theory, which are

\begin{equation}
[{\bf 1}_1 \oplus {\bf \overline {3}}_{1/2} \oplus {\bf 3}_0
\oplus {\bf 1}_{-1/2} ]
\oplus [{\bf 1}_{1/2} \oplus {\bf \overline {3}}_{0} \oplus {\bf 3}_{-1/2}
\oplus {\bf 1}_{-1}],
\end{equation}
in terms of two combined SU(3/1)$\otimes$U$^i$(1) complex reps
with $h^a=h^c= \pm \frac{7}{2}, h^e=h^g= \mp \frac{3}{2}$ such as

\begin{equation}
\begin{array}{lcll}
\mbox{floor}     & \mbox{SU(3/1)}     & \mbox{field} & \mbox{helicity} \\
\\
\mid \mbox{gnd}> & (0~0~-\frac{3}{2}) & A_\mu             & +1 \\
\\
\mid \mbox{1st}> & (0~1~-\frac{3}{2}) & \overline 3 \lambda & +\frac{1}{2} \\
\\
\mid \mbox{2nd}> & (1~0~-\frac{1}{2}) & 3 \phi              & ~0 \\
\\
\mid \mbox{3rd}> & (0~0~+\frac{1}{2}) & \chi        & -\frac{1}{2}, \\
\end{array}
\end{equation}

\noindent
and

\begin{equation}
\begin{array}{lcll}
\mbox{floor}     & \mbox{SU(3/1)}     & \mbox{field}     & \mbox{helicity} \\
\\
\mid \mbox{gnd}> & (0~0~-\frac{1}{2}) & \chi             & +\frac{1}{2} \\
\\
\mid \mbox{1st}> & (0~1~-\frac{1}{2}) & \overline 3 \phi  & ~0 \\
\\
\mid \mbox{2nd}> & (1~0~+\frac{1}{2}) & 3 \lambda         & -\frac{1}{2} \\
\\
\mid \mbox{3rd}> & (0~0~+\frac{3}{2}) & A_\mu              & -1.\\
\end{array}
\end{equation}

\vspace{1cm}
\begin{center}
{\bf 3.6 {\it N}=2 Supergravity}
\end{center}

Next, let us carry out superalgebraic truncation from the {\it N}=3
to the {\it N}=2
theory.  By using {\bf P$_2$}(3), we get {\it N}=2 multiplets from the
branching rule SU(3/1)$\rightarrow$ SU(2/1)$\otimes$U$^k$(1) as follows:

\begin{equation}
\begin{array}{ccl}

(0~0~-\frac{7}{2}) & \rightarrow & (0~-\frac{7}{2})(+\frac{7}{2})
                        \oplus(0~-\frac{5}{2})(+\frac{3}{2}), \\
\\
(0~0~+\frac{3}{2}) & \rightarrow & (0~+\frac{5}{2})(-\frac{7}{2})
                         \oplus(0~+\frac{3}{2})(-\frac{3}{2}). \\
\end{array}
\end{equation}
\\
\noindent
The typical rep $(0~-\frac{7}{2})$ with $h^a=h^c=h^e=h^g=h^i=h^k=
\frac{7}{2}$
is  $[{\bf 2}_B \oplus {\bf 2}_F]_\uparrow$, which
contains the half of the supermultiplets of the
{\it N}=2 theory as follows:

\begin{equation}
\begin{array}{lcll}
\mbox{floor}     & \mbox{SU(2/1)}     & \mbox{field}    & \mbox{helicity} \\
\\
\mid \mbox{gnd}> & (0~-\frac{7}{2})   & e_\mu^a         & +2 \\
\\
\mid \mbox{1st}> & (1~-\frac{7}{2})   & 2 \Psi_\mu      & +\frac{3}{2} \\
\\
\mid \mbox{2nd}> & (0~-\frac{5}{2})   & A_\mu           & +1 \\
\end{array}
\end{equation}

\noindent
where the helicity quantum number is given by

\begin{equation}
h = -\frac{1}{84} h^a -\frac{1}{60} h^c -\frac{1}{40} h^e
-\frac{1}{24} h^g -\frac{1}{12} h^i -\frac{1}{4} h^k -\frac{1}{2} h^l .
\end{equation}

\noindent
The complex conjugate rep $(0~+\frac{3}{2})$ with $h^a=h^c=h^e=h^g=h^i=h^k=
-\frac{7}{2}$ is $[{\bf 2}_B \oplus {\bf 2}_F]_\downarrow$,
which contains the
remaining half of the supermultiplets as follows:

\begin{equation}
\begin{array}{lcll}
\mbox{floor}     & \mbox{SU(2/1)}     & \mbox{field}    & \mbox{helicity} \\
\\
\mid \mbox{gnd}> & (0~+\frac{5}{2})   & A_\mu           & -1 \\
\\
\mid \mbox{2nd}> & (1~+\frac{5}{2})   & 2 \Psi_\mu      & -\frac{3}{2} \\
\\
\mid \mbox{3rd}> & (0~+\frac{7}{2})   & e_\mu^a         & -2. \\
\end{array}
\end{equation}
\\
As a result, the supergravity multiplets $[{\bf 4}_B \oplus {\bf 4}_F]$
of the {\it D}=4, {\it N}=2 theory, which is given by

\begin{equation}
[{\bf 1}_2 \oplus {\bf 2}_{3/2} \oplus {\bf 1}_1 ]
\oplus [{\bf 1}_{-1} \oplus {\bf 2}_{-3/2} \oplus {\bf 1}_{-2}],
\end{equation}

\noindent
are identified with the two irreps in Eqs.(49) and (51).

On the other hand, the Yang-Mills multiplet of {\it N}=3 splits into one
Yang-Mills multiplet and two matter multiplets. The Yang-Mills multiplet
$[{\bf 4}_B \oplus {\bf 4}_F]$ of {\it N} =2 theory, which consists of

\begin{equation}
[{\bf 1}_1 \oplus {\bf 2}_{1/2} \oplus {\bf 1}_0 ]
\oplus [{\bf 1}_{0} \oplus {\bf 2}_{-1/2} \oplus {\bf 1}_{-1}],
\end{equation}
in terms of two combined SU(2/1)$\otimes$U$^k$(1) complex reps
with $h^a=h^c= \pm \frac{7}{2}, h^e=h^g= \mp \frac{3}{2},
h^i=h^k= \pm \frac{3}{2}$, is given by

\begin{equation}
\begin{array}{lcll}
\mbox{floor}     & \mbox{SU(2/1)}     & \mbox{field}    & \mbox{helicity} \\
\\
\mid \mbox{gnd}> & (0~-\frac{3}{2})   & A_\mu           & +1 \\
\\
\mid \mbox{1st}> & (1~-\frac{3}{2})   & 2\lambda        & +\frac{1}{2} \\
\\
\mid \mbox{2nd}> & (0~-\frac{1}{2})   & \phi            & ~0 \\
\end{array}
\end{equation}

\noindent
and

\begin{equation}
\begin{array}{lcll}
\mbox{floor}     & \mbox{SU(2/1)}     & \mbox{field}    & \mbox{helicity} \\
\\
\mid \mbox{gnd}> & (0~+\frac{1}{2})   & \phi            & ~0 \\
\\
\mid \mbox{1st}> & (0~+\frac{1}{2})   & 2\lambda        & -\frac{1}{2} \\
\\
\mid \mbox{2nd}> & (0~+\frac{3}{2})   & A_\mu           & -1.\\
\end{array}
\end{equation}
\\
On the other hand, each matter multiplet $[{\bf 2}_B
\oplus {\bf 2}_F]$ of {\it N} =2
theory, which is

\begin{equation}
[{\bf 1}_{1/2} \oplus {\bf 2}_{0} \oplus {\bf 1}_{-1/2} ],
\end{equation}
in terms of SU(2/1)$\otimes$U$^k$(1) real rep
with $h^a=h^c= \pm \frac{7}{2}, h^e=h^g= \mp \frac{3}{2},
h^i=h^k= \pm \frac{3}{2}$,
is given by

\begin{equation}
\begin{array}{lcll}
\mbox{floor}     & \mbox{SU(2/1)}     & \mbox{field}    & \mbox{helicity} \\
\\
\mid \mbox{gnd}> & (0~-\frac{1}{2})   & \lambda_{+}     & +\frac{1}{2} \\
\\
\mid \mbox{1st}> & (1~-\frac{1}{2})   & 2\Phi           & ~0 \\
\\
\mid \mbox{2nd}> & (0~+\frac{1}{2})   & \lambda_{-}     & -\frac{1}{2}. \\
\end{array}
\end{equation}

\vspace{1cm}
\begin{center}
{\bf 3.7 {\it N}=1 Supergravity}
\end{center}

Finally, let us carry out superalgebraic truncation
from the {\it N}=2 to the
{\it N}=1 theory. In this case, although all gauge symmetries
are completely broken
except U(1) helicity, let us consider the {\it N} = 1 multiplets
comparing with
{\it N} = 2 rep for the sake of uniform notation. By using {\bf P$_2$}(2),
we get
{\it N}=1 multiplets from {\it N}=2 multiplets by using the branching rule
SU(2/1)(~$\supset$SU(2)$\otimes$U$^l$(1)~)
$\rightarrow$ U$^m$(1)$\otimes$U$^l$(1).
{}From the full weight system of the typical rep $(0~-\frac{7}{2})$ with
$h^a=h^c=h^e=h^g=h^i=h^k= \frac{7}{2}$,
one can obtain the halves of the graviton and extra gravitino
supermultiplets of the
{\it N}=1 theory, which are

\begin{equation}
[{\bf 1}_2 \oplus {\bf 1}_{3/2}]_\uparrow\oplus[{\bf 1}_{3/2}
\oplus {\bf 1}_1],
\end{equation}
as follows:

\begin{equation}
\begin{array}{lccll}
\mbox{floor}     & \mbox{SU(2/1)}     & \mbox{U$^m$(1)$\otimes$U$^l$(1)}
& \mbox{field}      & \mbox{helicity} \\
\\
\mid \mbox{gnd}> & (0~-\frac{7}{2})   & (+\frac{7}{2})(-7)
& e_\mu^a           & +2 \\
\\
\mid \mbox{1st}> & (1~-\frac{7}{2})   & (+\frac{7}{2})(-6)
& \Psi_\mu          & +\frac{3}{2} \\
\\
                 & (-1~-\frac{5}{2})  & (+\frac{5}{2})(-6)
                 & \chi_\mu          & +\frac{3}{2} \\
\\
\mid \mbox{2nd}> & (0~-\frac{5}{2})   & (+\frac{5}{2})(-5)
& A_\mu             & +1, \\
\end{array}
\end{equation}
\\
\noindent
where the helicity quantum number {\it h} is the same form as that of Eq.(50).
Although $h^m$ quantum number do not contribute to the helicity,
it is important
to classify the supermultiplets.  From the complex conjugate rep
$(0~+\frac{3}{2})$
with $h^a=h^c=h^e=h^g=h^i=h^k= -\frac{7}{2}$, we obtain

\begin{equation}
[{\bf 1}_{-1} \oplus {\bf 1}_{-3/2}]_\downarrow\oplus[{\bf 1}_{-3/2}
\oplus {\bf 1}_{-2}],
\end{equation}
which are the remaining halves of the supermultiplets as follows:

\begin{equation}
\begin{array}{lccll}
\mbox{floor}     & \mbox{SU(2/1)}     & \mbox{U$^m$(1)$\otimes$U$^l$(1)}
& \mbox{field}      & \mbox{helicity} \\
\\
\mid \mbox{gnd}> & (0~+\frac{5}{2})   & (-\frac{5}{2})(5)
& A_\mu             & -1 \\
\\
\mid \mbox{1st}> & (1~+\frac{5}{2})   & (-\frac{5}{2})(6)
& \chi_\mu          & -\frac{3}{2} \\
\\
                 & (-1~+\frac{7}{2})  & (-\frac{7}{2})(6)
                 & \Psi_\mu          & -\frac{3}{2} \\
\\
\mid \mbox{2nd}> & (0~+\frac{7}{2})   & (-\frac{7}{2})(7)
& e_\mu^a           & -2. \\
\end{array}
\end{equation}

As a result, the desired graviton multiplet $[{\bf 2}_B \oplus {\bf 2}_F]$
of the {\it D}=4, {\it N}=1 theory, which is given by

\begin{equation}
[{\bf 1}_2 \oplus {\bf 1}_{3/2} ]
\oplus [{\bf 1}_{-3/2} \oplus {\bf 1}_{-2}],
\end{equation}

\noindent
is obtained from two irreps in Eqs.(59) and (61)
by removing the extra gravitino part.

On the other hand, the Yang-Mills multiplet  of {\it N}=2 splits
into Yang-Mills
and matter multiplets of {\it N}=1 theory. As a result, we have the
Yang-Mills multiplet $[{\bf 2}_B
\oplus {\bf 2}_F]$ of {\it N} =1 theory, which is

\begin{equation}
[{\bf 1}_1 \oplus {\bf 1}_{1/2} ]
\oplus [{\bf 1}_{-1/2} \oplus {\bf 1}_{-1}],
\end{equation}
and the matter multiplet $[{\bf 2}_B \oplus {\bf 2}_F]$, which is

\begin{equation}
[{\bf 1}_{1/2} \oplus {\bf 1}_{0} ]
\oplus [{\bf 1}_{0} \oplus {\bf 1}_{-1/2}],
\end{equation}
in terms of the full weight system of two combined SU(2/1) complex
reps of {\it N}=2 theory
with $h^a=h^c= \pm \frac{7}{2}, h^e=h^g= \mp \frac{3}{2},
h^i=h^k= \pm \frac{3}{2}$
as follows

\begin{equation}
\begin{array}{lccll}
\mbox{floor}     & \mbox{SU(2/1)}     & \mbox{U$^m$(1)$\otimes$U$^l$(1)}
& \mbox{field}      & \mbox{helicity} \\
\\
\mid \mbox{gnd}> & (0~-\frac{3}{2})   & (\frac{3}{2})(-3) & A_\mu    & +1 \\
\\
\mid \mbox{1st}> & (1~-\frac{3}{2})   & (\frac{3}{2})(-2) & \lambda
& +\frac{1}{2} \\
\\
 & (1~+\frac{1}{2})   & (\frac{1}{2})(-2) & \chi        & +\frac{1}{2} \\
\\
\mid \mbox{2nd}> & (0~-\frac{1}{2})   & (\frac{1}{2})(-1) & \phi  & ~0, \\
\end{array}
\end{equation}

\noindent
and

\begin{equation}
\begin{array}{lccll}
\mbox{floor}     & \mbox{SU(2/1)} & \mbox{U$^m$(1)$\otimes$U$^l$(1)}&
\mbox{field} & \mbox{helicity} \\
\\
\mid \mbox{gnd}> & (0~+\frac{1}{2})   & (-\frac{1}{2})(1) & \phi
& ~0 \\
\\
\mid \mbox{1st}> & (1~+\frac{1}{2})   & (-\frac{1}{2})(2) & \chi
& -\frac{1}{2} \\
\\
                 & (-1~+\frac{3}{2})   & (-\frac{3}{2})(2) & \lambda
                 & -\frac{1}{2} \\
\\
\mid \mbox{2nd}> & (0~+\frac{3}{2})   & (-\frac{3}{2})(3) & A_\mu
& -1.\\
\end{array}
\end{equation}

On the other hand, we can also obtain the matter multiplet
$[{\bf 2}_B \oplus {\bf 2}_F]$ of {\it N} =1
theory from the matter multiplet of {\it N}=2 theory, which is

\begin{equation}
[{\bf 1}_{1/2} \oplus {\bf 2}_{0} \oplus {\bf 1}_{-1/2}],
\end{equation}

\noindent
in terms of the full weight system of SU(2/1) real rep
with $h^a=h^c= \pm \frac{7}{2}, h^e=h^g= \mp \frac{3}{2},
h^i=h^k= \pm \frac{3}{2}$
as follows

\begin{equation}
\begin{array}{lccll}
\mbox{floor}     & \mbox{SU(2/1)}     & \mbox{U$^m$(1)$\otimes$U$^l$(1)}
& \mbox{field}      & \mbox{helicity} \\
\\
\mid \mbox{gnd}> & (0~-\frac{1}{2})   & (+\frac{1}{2})(-1) & \lambda_{+}
& +\frac{1}{2} \\
\\
\mid \mbox{1st}> & (1~-\frac{1}{2})   & (+\frac{1}{2})(0) & \Phi  & ~0 \\
\\
                 & (-1~+\frac{1}{2})  & (-\frac{1}{2})(0) & \Phi  & ~0 \\
\\
\mid \mbox{2nd}> & (0~+\frac{1}{2})   & (-\frac{1}{2})(1) & \lambda_{-}
& -\frac{1}{2}. \\
\end{array}
\end{equation}

\vspace{2cm}

\begin{center}
{\large \bf IV. Conclusion}
\end{center}

In conclusion, we have studied {\it D}=4, {\it N}=8 supergravity
in the context
of SU(8/1) Lie superalgebra. We have constructed suitable general projection
matrix, and used this matrix to obtain all possible regular maximal branching
patterns in terms of Kac-Dynkin weight techniques.  Then, we have shown that
consistent successive superalgebraic trunctions from the {\it D}=4, {\it N}=8
maximal theory to the {\it D}=4, {\it N}=7, 6, ${\cdots}$, 1 theories can be
systematically realized as sub-algebra chains of the SU(8/1) superalgebra.
As results, we have identified all possible supermultiplets of {\it N}=8, 7,
${\cdots}$, 1 theories, which have been classified in terms of
super-Poincar\'{e}
algebra by Strathdee, with irreps of SU({\it N}/1) superalgebra.  Finally, we
hope that our branching technique will provide a deeper understanding of
the structure of the supersymmetric systems and help finding new
supersymmetric theories.

\vspace{1cm}

\begin{center}
{\bf ACKNOWLEDGEMENTS}
\end{center}

The present study was supported by the Basic Science Research
Institute Program,
Ministry of Education, 1994, Project No. 2414.

\newpage

\begin{center}
{\bf REFERENCES}
\end{center}

\begin{description}
\item{[1]} V. Kac, Adv. in Math. {\bf 26}, 8 (1977); Commun. Math. Phys.
{\bf 53}, 31 (1977).
\item{[2]} Y.A. Gol'fand and E.P. Likhtman, Pis'ma Zh. Eksp. Theor. Fiz.
 {\bf 13}, 452(1971)[JETP Lett. {\bf 13},323(1971)]; P. Ramond, Phys. Rev.
 {\bf D3}, 2415 (1971); P. G. O. Freund and  I. Kaplansky, J. Math. Phys.
 {\bf 17}, 228 (1976); J. P. Hurni and B. Morel,  {\it ibid.} {\bf 24},
 157 (1983).
\item{[3]} Y. Ne'eman, Phys. Lett. {\bf 81B}, 190 (1979); D. B. Fairlie,
 {\it ibid.} {\bf 82B}, 97 (1979).
\item{[4]} F. Iachello, Phys. Rev. Lett. {\bf 44}, 772 (1980);
 A.B. Balantekin,I. Bars, and F. Iachello, {\it ibid.} {\bf 47}, 19 (1981).
\item{[5]} S. Deser and B. Zumino, Phys. Lett. {\bf 62B}, 335 (1976);
D. Z. Freedman, P. van Nieuwenhuizen, and S. Ferrara, Phys. Rev. {\bf D13},
3214 (1976); P. van Nieuwenhuizen, Phys. Rep. {\bf 68}, 189 (1981).
\item{[6]} A. Neveu and J. H. Schwarz, Nucl. Phys. {\bf B31}, 86 (1971);
M. B. Green and J. H. Schwarz, {\it ibid.} {\bf B198}, 474 (1982).
\item{[7]} J. Strathdee, Int. J. Mod. Phys. {\bf A2}, 173 (1987).
\item{[8]} E. Cremmer, in {\it Superspace and Supergravity}, edited by
S. Hawking and M. Rocek (Cambridge University Press, London, England, 1980).
\item{[9]} K. Y. Kim, C. H. Kim, Y. Kim, and Y. J. Park,
J. Korean Phys. Soc. {\bf 18}, 249 (1985);  C. H. Kim,
K. Y. Kim, W. S. l'Yi, Y. Kim, and Y. J. Park, Mod. Phys. Lett. {\bf A3},
1005 (1988); C. H. Kim, Y. J. Park, K. Y. Kim, Y. Kim, and W. S. l'Yi, Phys.
Rev. {\bf D44}, 3169 (1991).
\item{[10]}C. H. Kim, K. Y. Kim, Y. Kim, H. W. Lee, W. S. l'Yi, and
Y. J. Park, Phys. Rev. {\bf D40}, 1969 (1989).
\item{[11]} C. H. Kim, K. Y. Kim, W. S. l'Yi, Y. Kim, and Y. J. Park,
J. Math. Phys. {\bf 27}, 2009 (1986).
\item{[12]} M. B. Green and J. H. Schwarz, Phys. Lett. {\bf 122B}, 143 (1983).
\item{[13]} J. H. Schwarz and P. C. West, Phys. Lett. {\bf 126B}, 301 (1983);
P. Howe and P. C. West, Nucl. Phys. {\bf B238}, 181 (1984).
\item{[14]} C. H. Kim, K. Y. Kim, Y. Kim, and Y. J. Park, Phys.
Rev. {\bf D39}, 2967 (1989); C. H. Kim, S. Cho, S. H. Yoon, and Y. J. Park,
J. Korean Phys. Soc. {\bf 26}, 603 (1993).
\item{[15]} R. Slansky, Phys. Rep. {\bf 79}, 1 (1981); C. H. Kim, Y. J. Park,
I. G. Koh, K. Y. Kim, and Y. Kim, Phys. Rev. {\bf D27}, 1932 (1983).
\item{[16]} C. H. Kim, Y. J. Park, K. Y. Kim, and Y. Kim,
J. Korean Phys. Soc. {\bf 25}, 87 (1992).

\end{description}
\end{document}